# Emerging Connections: Quantum and Classical Optics

Xiao-Feng Qian, A. Nick Vamivakas and Joseph H. Eberly

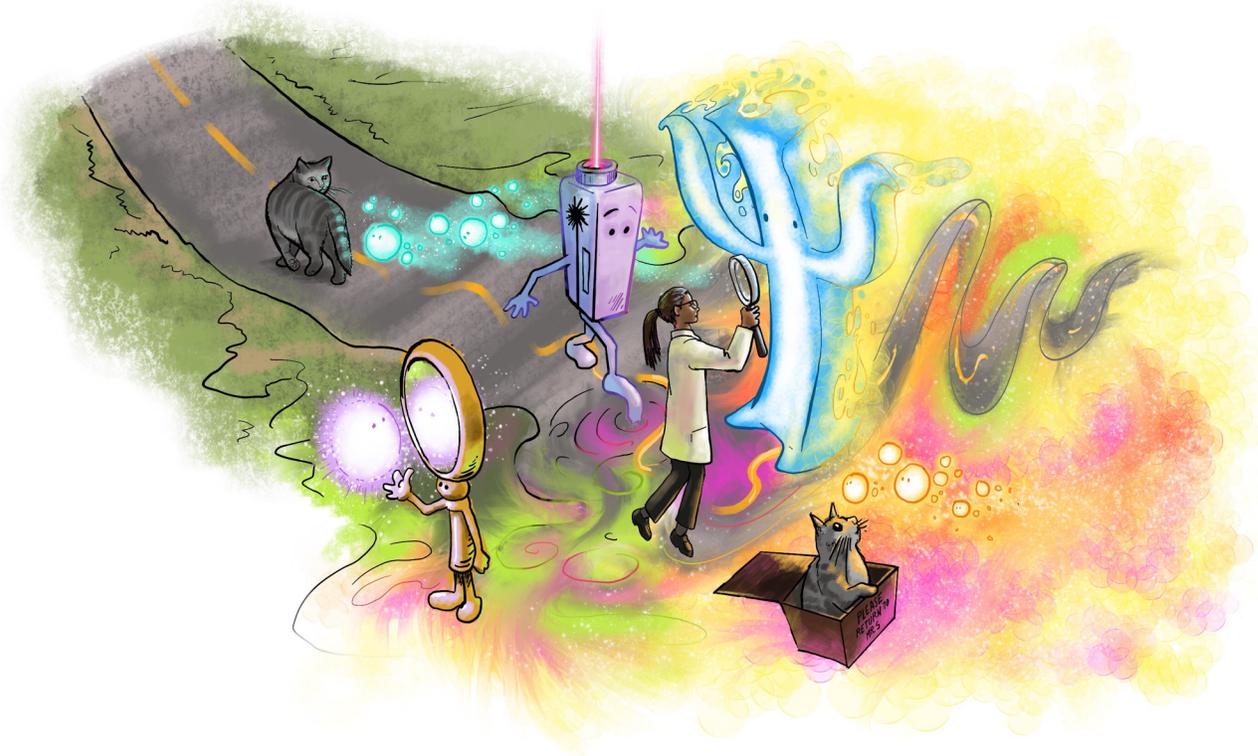

The blurring of the classical-quantum boundary points to new directions in optics

Quantum optics and classical optics have coexisted for nearly a century as two distinct, self-consistent descriptions of light. What influences there were between the two domains all tended to go in one direction, as concepts from classical optics were incorporated into quantum theory's early development. But it's becoming increasingly clear that a significant quantum presence exists in classical territory—and, in particular, that the quintessential quantum attribute, entanglement, can be seen, studied and exploited in classical optics.

This blurring of the classical-quantum boundary has opened up a potential new direction for frontier work in optics. In recognition of that, an OPN Incubator Meeting in Washington, D.C., in November 2016 focused on the emerging connections between the classical and quantum approaches to optics, and where the field is going. In this article we overview some of the topics that informed the Incubator.

**Entanglement and optics**

Entanglement first became a topic of wide interest to the optics community because of its importance to quantum information. In a 1998 study, however, Robert J.C. Spreeuw argued that entanglement is compatible with classical wave optics, an opening that has subsequently received careful analysis from many authors and groups.

The very notion of "classical entanglement" is challenging; no less a figure than Erwin Schrödinger, after all, in a 1935 analysis of the status of quantum mechanics, called entanglement "not … *one* but rather *the* characteristic trait of quantum mechanics" (the italics are Schrödinger's). Today many optical applications are dependent on quantum mechanics, optical physics includes only a few uniquely quantum processes, all of them identified with single-photon actions—such as photon detection, photon counting and spontaneous emission—for which classical counterparts don't exist.

Yet Schrödinger himself likely knew that entanglement was not unique to quantum mechanics—a fact suggested by a reference in the same 1935 paper to the functional-analysis theorem developed by Erhard Schmidt in 1907, which is frequently used today in both quantum and classical entanglement theory. In addition to entanglement, the widely accepted role of so-called Bell-inequality violations as a marker of true quantum character has prompted tests using entirely classical light, without photon registration. The success reported for those tests raises additional issues, such as the role of Einsteinian locality and the importance of vector-space analysis engaging the Schmidt theorem across the classical-quantum boundary.

**Spin, polarization and entanglement**

All of those issues stimulated vigorous discussions at the November 2016 Incubator—an event that also exposed a growing number of roles for entanglement in addressing questions in classical optics. Those expanding roles stem, in part, from some useful aspects of the same Schmidt theorem cited by Schrödinger in 1935.

One form of quantum-like classical entanglement involves the intrinsic angular momentum of a light field, or its optical spin—which can be thought of as equivalent to ordinary light polarization. A standard expression relating the amplitudes of a transverse optical beam field, $E$, its spin (polarization) vectors, $x$ and $y$, is

$$\vec{E}(\vec{r}, t) = \hat{x} E_x(\vec{r}, t) + \hat{y} E_y(\vec{r}, t). \quad (1)$$

Though the fact is commonly overlooked, this equation actually shows amplitude-spin entanglement: the optical field $E$ is just a sum of the products of the spin degree of freedom and the amplitude degree of freedom. The Schmidt theorem lends a hand here, because it holds that a continuously infinite-dimensional function space, such as amplitude, becomes effectively only two-dimensional when in an entangled combination with a two-dimensional degree of freedom, such as spin.

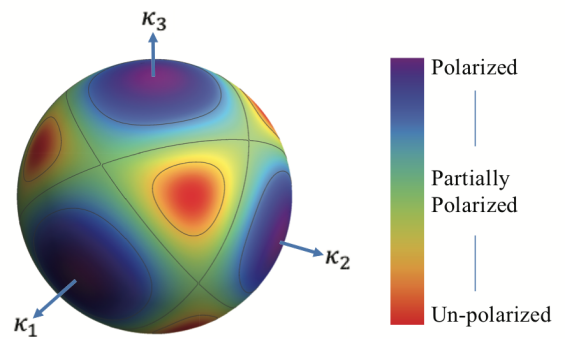

**Polarization:** The degree of polarization defined as the opposite if entanglement for a point-like electromagnetic field by Qian and Eberly, Opt. Lett. 36, 4110 (2011).

Using quantum language, this means that the equation relating spin and amplitude can be analyzed as an entangled state of two qubits—the simplest entanglement known. That's true even though the two quantities are vectors in very different Hilbert spaces and are clearly non-quantum in character. (For more on the mathematical formalism of classical entanglement, see sidebar, "Classical entanglement"). Since optical fields are supplied with independent spatial and temporal degrees of freedom, such entanglement is available in a number of forms. The opening plenary remarks by Francisco De Zela and Wolfgang Schleich at the November 2016 OSA Incubator spurred discussions of many related aspects.

**Classical Entanglement:** A key to classical entanglement lies in the conditional logic of interpreting multi-wave superposition. Consider a wave signal depending on temporal modes $F_k(t)$ (referring, say, to different wavelengths) and spatial modes $G_m(r)$. Two different versions of that signal employ superpositions of two temporal modes and two spatial modes:

$$V_A(r,t) = [F_1(t) + F_2(t)] \otimes [G_1(r) + G_2(r)] \quad \text{and} \quad V_B(r,t) = F_1(t) \otimes G_1(r) + F_2(t) \otimes G_2(r)$$

where the $\otimes$ symbol indicates a tensor product—necessary because $F$ and $G$ belong to different vector spaces—and where the $+$ symbol takes care of summation of vectors within the same vector space (ordinary functional superposition). We can recast this expression in the case of a primitive "measurement" of red and blue wavelengths separated by a prism (as shown in the figure below, with the spatial-mode behaviors chosen to be the upper or lower half of a circle) as:

$$V_A = [\text{Red} + \text{Blue}] \otimes [\text{Upper} + \text{Lower}] \quad \text{and} \quad V_B = \text{Red} \otimes \text{Upper} + \text{Blue} \otimes \text{Lower}$$

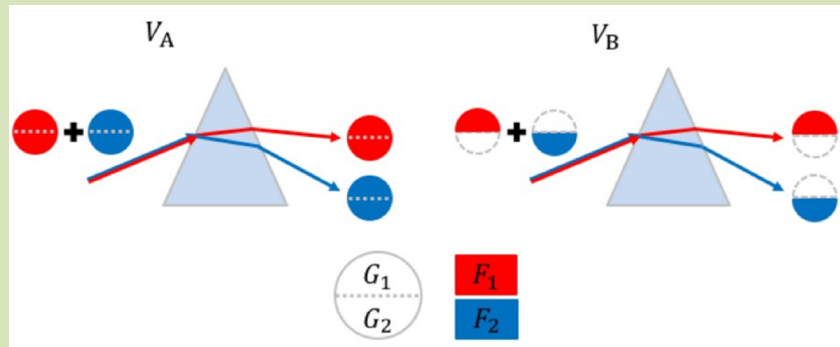

In the case of signal $V_B$, the detection of Blue conveys conditional information—namely, that spatial mode Upper is present but Lower is not. In the case of signal $V_A$, detection of Blue conveys no such spatial-mode information, since both Upper and Lower accompany Blue. Another way to express this is that signal $V_A$ is fully factored between wavelength and space mode, whereas signal $V_B$ is not factorable into a single product of wavelength and spatial modes. Because of the non-factorability or non-separability between the degrees of freedom—wavelength and spatial mode in this example—$V_B$ is called an entangled signal. (The spin and amplitude components discussed in the main text are analogs of these $F(t)$ and $G(r)$ modes.)

This example, tied to the factorability of two independent degrees of freedom is a *vector-space question*, not a quantum-classical distinction. Entanglement is defined for the Hilbert spaces in quantum and classical theories in the same way.

But what about when one of the two amplitude components is negligible—for example, when $|E_y| \ll |E_x|$, and the field is almost completely $x$-polarized, and $E \simeq xE_x$? In such a situation, the spin and amplitude degrees of freedom of $E$ are almost perfectly factored—not entangled. Thus, complete polarization is the same as zero entanglement, and complete entanglement implies zero polarization. Spin polarization and spin entanglement are opposites.

This insight has proved useful in eliminating a running controversy in some areas of practical importance, such as wide-aperture microscopy or hohlraum fields. In such situations, all three field components, $x$, $y$, and $z$, must be included for a correct description—but the traditional methods to obtain the numerical degree of polarization are ambiguously different. Schmidt's theorem, however, once again removes the roadblock: it lets one determine the degree of polarization via its reverse, the degree of entanglement, because Schmidt's theorem shows that the degree of entanglement is unambiguous for any number of field components.

### Optical metrology and quantum information

Quantum insights are empowering classical optical technology in other ways as well—particularly through strong correlations in multiple-degree-of-freedom (DoF) optical beams.

Polarimetry can use entanglement between spin polarization and the modes that describe a beam's transverse spatial structure. A beam can arrive at a target with multiple polarization states simultaneously; measurements based on entanglement of the polarization states with the spatial modes can be used to recover the polarization information. The end result is a metrology tool explored by Andrea Aiello and colleagues from the Max Planck Institute, Germany, to perform single-shot Mueller matrix tomography. Remarkably, in the context of polarimetry, a team led by B.N. Simon of IIT Madras, India, has shown that classical entanglement can be used to resolve a longstanding question about the subset of matrices that are physically reasonable Mueller matrices.

Polarization–spatial-mode entanglement is not limited to polarimetry. It can also provide a means to monitor the real-time kinematics of an illuminated target particle, and to locate the position of a particle within the beam with a detector-limited temporal resolution. (Such a feat was recently demonstrated by the research group of Gerd Leuchs at the Max Planck Institute, work shared in a talk by Aiello at the Incubator.) The entanglement allows the particle position to become encoded in the polarization DoF, recovered via polarization tomography. Many future novel metrology and sensing approaches will likely take advantage of the parallelism offered by classically entangled beams.

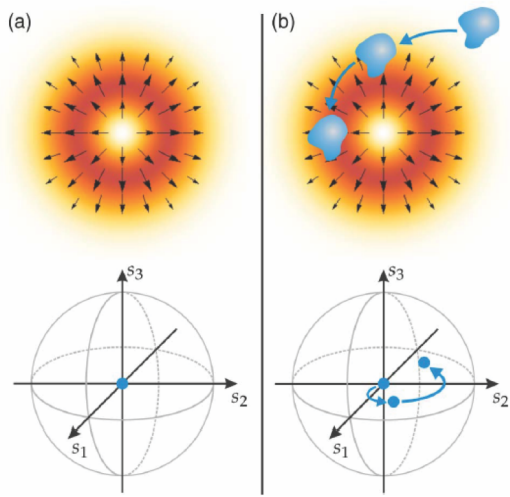

**Metrology:** Optical sensing and metrology by Berg-Johansen, et al., Optica, 2, 864 (2015).

Also intriguing is the prospect of leveraging quantum-like correlations, such as entanglement in classical beams, for tasks associated with quantum information. For example, entangled classical correlations can act as a channel for some quantum-like communication tasks. One such task is quantum teleportation, which allows the transfer of quantum states among independent parties or independent degrees of freedom, particles or waves. Demonstrations by the research groups of Robert Boyd, University of Rochester, USA, and of Antonio Khoury, Universidade Federal Fluminense, Brazil, have already shown that it is possible to teleport information between orbital angular momentum spatial modes and spin polarization modes.

## Addressing Fundamental Issues

Quantum insights have provided a more nuanced understanding of the nature of classical light, engaging issues as primary as the Bell inequality and changing perspectives on a fundamental property, coherence.

### *Bell-inequality violation*

The Bell inequality and its violation starred in a number of discussions in the Incubator, touching on both quantum and classical perspectives. The Incubator participants generally acknowledged that entanglement is the "secret ingredient" in Bell-inequality violation: entanglement is available in quantum theory and in classical wave optics as well—and no Bell violation has been achieved without it.

> **Bell Inequalities and Locality:** In the famous 1935 Einstein-Podolsky-Rosen (EPR) paper, Albert Einstein, with colleagues Boris Podolsky and Nathan Rosen, imagined a scenario in which two no longer interacting quantum systems described by a common (entangled) state are located extremely remotely from each other. That setup, sometimes called Einsteinian nonlocality, was to make it clear that measurements posited by EPR on the two systems would obviously be independent. Because quantum entanglement would require instantaneous interaction of these widely separated particles, Einstein concluded that quantum theory must be considered incomplete.
>
> Thirty years later, John Bell devised the first practical way to approach EPR experimentally by engaging electron spin (or, equivalently, light polarization). Bell showed that the average of the product $AB$ of two spin-like independent variables $A$ and $B$ that randomly take the values +1 or −1 could not possibly be made larger than a specific limit. Even if the supposedly random ±1 values are under the control of ideally designed non-quantum "hidden" variables, a mathematical constraint, now known as a Bell inequality places an upper limit of 2 on measure of their correlation, $S$, called the Bell measure.
>
> However, in the 1970s and 1980s, groups led by John Clauser and, later, Alain Aspect showed that if the ±1 values are obtained experimentally from recorded polarization projections via individual photon detections, and the average of $AB$ is evaluated for specific photonic polarization states, values strongly violating $S \leq 2$ are obtained. The result is what is now called a Bell violation ($S > 2$), generally considered a signpost of the quantum-classical boundary.
>
> 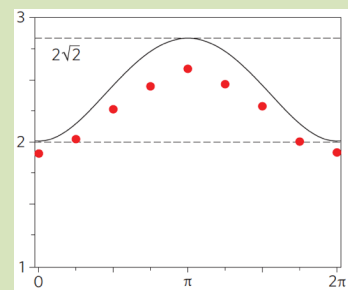
>
> K.H. Kagalwala et al., Nat. Photon. **7**, 72 (2013).
>
> However, recently since around 2010, a number of groups have experimentally registered Bell violation by fully classically-entangled optical beams (see for example the above red-dot data of S by Bahaa E. A. Saleh's group), questioning the validity of Bell violation as a quantum-classical marker.

A Bell-inequality violation has generally been viewed the marker that one has left classical physics and entered the quantum domain. But major advance came in 2010 when Khoury's research group achieved a Bell violation using classically entangled laser light. Bell violation has since been achieved under a number of classical conditions, including an experiment by Joseph Eberly's group at Rochester that employed strong thermal light from a below-threshold laser diode, a power meter for detection and almost perfectly unpolarized light for randomness. In a variety of cases, violation by many standard deviations has been achieved classically.

*Correlation, coherence and independent DoFs*
To cohere is to "stick together," or be united. In optics this can be applied to the various DoFs that are available. For example, if all or almost all modes of a light field are associated with a single value of another DoF—say, with the vertical spin orientation—then the field is said to have spin coherence, in the sense of being strongly vertically polarized. The recent, surprising discovery of several previously unrecognized coherences, by manipulation of degrees of freedom, has marked out a new avenue for research, and points to several unexpected applications.

|   | S | R | T |
|---|---|---|---|
| S | $s_1$-$s_2$ | s-r (?) | s-t (?) |
| R | r-s (?) | $r_1$-$r_2$ | r-t (?) |
| T | t-s (?) | t-r (?) | $t_1$-$t_2$ |

**Correlations:** The labels S, R, T stand for three degrees of freedom: spin, space, and time on which the field depends. The off-diagonal question marks indicate that, at least conventionally, different degrees of freedom that are independent cannot be correlated. But the presence of entanglement changes the rules of the correlation game.

The coherences of an optical field are measured by the strength of correlation functions. Strong temporal coherence of a signal $V$, for example, means that its auto-correlation function, $C(t_1 - t_2) = \langle V^*(t_2)V(t_1)\rangle$, remains nonzero even for large displacements $|t_1 - t_2|$. The same consideration applies to all DoFs, and their usual correlation functions can be said to be "diagonal," in the sense of the matrix in the figure at left/at right/above/below, in which the temporal correlation category occupies the lower-right corner.

The correlations suggested by the off-diagonal elements of the matrix are unusual, and at first sight impossible. By definition, independent degrees of freedom have no joint correlation—for example, spin has no influence on either the spatial or temporal features of a field, and is thus independent of them and uncorrelated with them. However, the existence of entanglement allows a very useful violation of this obvious truth.

*Hidden coherence*
The tension between polarization and entanglement described earlier—that is, the fact that obtaining more of either one means surrendering some of the other—puts them on a parallel footing; entanglement and polarization are alternative forms of coherence. As several speakers at the November 2016 Incubator emphasized, these are defined by pairs of DoFs rather than by single ones.

For example, projecting $E$ on the temporal modes allows space-spin coherence to emerge, which will obviously be different from the time-spin coherence obtained by projecting $E$ on the spatial modes. Both are polarizations in the traditional spin sense, but are different and have now been measured as distinct by Abouraddy and colleagues in Florida.

Indeed, the very action of projecting can lead to a third, hidden coherence—a "polarization" without polarization. That is, projecting on spin leaves temporal and spatial modes to cohere and entangle, providing an opportunity to obtain "polarization" in a completely new, non-spin sense, as shown recently by Xiao-Feng Qian and colleagues. All of these coherences identified by projection are associated with distinct two-party entanglements—space-spin, time-spin, and time-space—and all satisfy the same relation:

$$C_{ab}^2 + P_{ab}^2 = 1, \qquad (2)$$

where $a$ and $b$ refer to the DoF pair that is polarized/entangled, and the concurrence, $C$, measures the degree of entanglement.

This is not the end of exploration for hidden coherences. By failing to make any projection, the electric field, described earlier as an entanglement of spin and amplitude, becomes the strict analog of a three-party entangled state in quantum theory—a superposition of products of vectors from three different vector spaces or DoFs. Superpositions of three-way products identify an entirely new category of coherence, exploration of which has just begun.

**Looking ahead**
Future work at the classical-quantum boundary will likely explore what role hyper-entanglement plays in classical optics, and how to think about mixed classical states. The deeper understanding explored in the November 2016 OSA Incubator meeting will certainly promote new, and exploitable, technological perspectives. It is becoming possible to implement quantum-like classical optical technologies that leverage the parallelism of the quantum world, using lab platform—classical optical beams—that is friendly, robust and easy to control. Quantum-inspired approaches in classical optics will improve the performance of future metrology, communication and imaging systems.

The emerging links between quantum and classical optics thus open a frontier and a new framework for dealing with classical light, and provide a guide for investigations to uncover a deeper understanding about the light's fundamental nature. Remarkably, a century and a half after the publication of Maxwell's equations, classical optical fields still have some surprises to reveal.

Xiao-Feng Qian (xiaofeng.qian@rochester.edu), A. Nick Vamivakas (nick.vamivakas@rochester.edu), and Joseph H. Eberly (eberly@pas.rochester.edu). Center for Coherence and Quantum Optics, Department of Physics & Astronomy, The Institute of Optics, University of Rochester, Rochester, NY 14627, USA